\documentclass[twocolumn,showpacs,preprintnumbers,amsmath,amssymb,prb,aps]{revtex4-1}
\usepackage{epsfig}
\usepackage{epstopdf}
\usepackage{graphicx}
\usepackage{dcolumn}
\usepackage{bm}
\usepackage{textcomp}
\usepackage{array}
\usepackage{subcaption}
\usepackage{booktabs}

\begin{document}

\title{Exploring the possibility of enhancing the \textit{figure-of-merit} ( $>$ 2) of Na$_{0.74}$CoO$_{2}$: A combined experimental and theoretical study}
\author{Shamim Sk$^{1,}$}
\altaffiliation{Electronic mail: shamimsk20@gmail.com}
\author{Jayashree Pati$^{2}$}
\author{R. S. Dhaka$^{2}$}
\author{Sudhir K. Pandey$^{3,}$}
\altaffiliation{Electronic mail: sudhir@iitmandi.ac.in}
\affiliation{$^{1}$School of Basic Sciences, Indian Institute of Technology Mandi, Kamand - 175005, India}
\affiliation{$^{2}$Department of Physics, Indian Institute of Technology Delhi, Hauz Khas, New Delhi - 110016, India}
\affiliation{$^{3}$School of Engineering, Indian Institute of Technology Mandi, Kamand - 175005, India}

\date{\today} 

\begin{abstract}

Search of new thermoelectric (TE) materials with high \textit{figure-of-merit} (ZT) is always inspired the researcher in TE field. Here, we present a combined experimental and theoretical study of TE properties of Na$_{0.74}$CoO$_{2}$ compound in high temperature region. The experimental Seebeck coefficient (S) is found to vary from 64 to 118 $\mu$V/K in the temperature range $300-620$ K. The positive values of S are indicating the dominating p-type behaviour of the compound. The observed value of thermal conductivity ($\kappa$) is $\sim$ 2.2 W/m-K at 300 K. In the temperature region $300-430$ K, the value of $\kappa$ increases up to $\sim$ 2.6 W/m-K and then decreases slowly till 620 K with the corresponding value of $\sim$ 2.4 W/m-K. To understand the experimental transport properties, we have carried out the theoretical calculations using spin-polarized and spin-unpolarized DFT and DFT+\textit{U} methods. The best matching between experimental and calculated values are observed when the spin-polarized calculation is done by chosen \textit{U} = 4 eV in DFT+\textit{U}. By taking calculated S and electrical conductivity ($\sigma$) along with experimental $\kappa$, we have optimized the ZT values up to 1200 K and the maximum value is found to be $\sim$ 0.67 at 1200 K. Our computational study suggests that the possibility of n-type behaviour of the compound which can lead to a large value of ZT at higher temperature region. Electron doping of $\sim$ 5.1$\times$10$^{20}$ cm$^{-3}$ is expected to give rise the high ZT value of $\sim$ 2.7 at 1200 K. Using these temperature dependent ZT values, we have calculated the maximum possible values of efficiency ($\eta$) of thermoelectric generator (TEG) made by p and n-type Na$_{0.74}$CoO$_{2}$. The present study suggests that one can get the efficiency of a TE cell as high as $\sim$ 11$\%$ when the cold and hot end temperature are fixed at 300 K and 1200 K, respectively. Such high values of ZT and efficiency suggest that Na$_{0.74}$CoO$_{2}$ can be used as a potential candidate for high temperature TE applications.           
   
\vspace{0.2cm}
\end{abstract}

\maketitle

\section{Introduction} 
Thermoelectric (TE) materials convert the waste heat into useful electricity\cite{electricity1,electricity2} and are believed to play an important role in the future new energy-conversion system. The efficiency of TE materials is evaluated by a dimensionless parameter, \textit{figure-of-merit} (ZT)\cite{zt1,zt2} 
\begin{equation}
ZT = \frac{S^{2}\sigma T}{\kappa}
\end{equation}
where, S is Seebeck coefficient, $\sigma$ is electrical conductivity, T is absolute temperature and $\kappa(=\kappa_{e} + \kappa_{ph})$ is thermal conductivity. For a good TE material, the value of ZT should be $\geq$ 1.\cite{snyder} Therefore, for efficient TE materials we need to have high S and $\sigma$ with low $\kappa$. But, the realization of high ZT value is difficult, because all these three parameters (S, $\sigma$ and $\kappa$) are interrelated to each other.\cite{shamim,ashcroft} From the last few decades, many experimental and theoretical approaches have been used to improve the value of ZT.\cite{shen,sakurada,biswas}

Till now, the best commercially used TE materials available are bismuth (and lead) telluride and its alloys with highest ZT $\sim$ 1 near the room temperature.\cite{bismuth,lead} However, some of these materials are rare, costly and toxic. In addition, these materials can't sustain at high temperature region. In this context, oxide materials have gained more attention in TE family due to their high temperature stability, low-cost, non-toxic, oxidation free and simple synthesis procedure. Na$_{x}$CoO$_{2}$ is one of the most studied TE candidate among the oxide materials in last two decades. The study of TE properties in Na$_{x}$CoO$_{2}$ was started when Terasaki \textit{et al.}\cite{tarasaki} reported a large thermopower (100 $\mu$V/K at 300 K) in NaCo$_{2}$O$_{4}$ in 1997. The spin entropy has been found to play an important role for this unusual large values of S in this system.\cite{wang} Na-site substitution effects on the thermoelectric properties of NaCo$_{2}$O$_{4}$ was studied by Kawata \textit{et al.}\cite{kawata}. The attention in this compound was again stimulated by the discovery of superconductivity at 5 K in Na$_{0.35}$CoO$_{2}$:1.3H$_{2}$O.\cite{superconductivity} Further investigation of the S in Na$_{x}$CoO$_{2}$ has reported by Kaurav \textit{et al.}\cite{kaurav} by varying the $x$ values from 0.23 to 0.84.

A detailed TE study on Na$_{x}$CoO$_{2}$ has been carried out by many groups\cite{tarasaki,kawata,kaurav,rivadulla, poor,vitta, ito, park} from last two decades. But, there are no such systematic experimental and theoretical study on TE properties of this compound available in high temperature region. Therefore, investigation of TE properties in this compound is required at elevated temperature. 

Na$_{x}$CoO$_{2}$ ($0.25<x\lesssim0.8$) possesse hexagonal symmetry with space group of 194; P63/mmc with two triangular CoO$_{2}$ layers per unit cell, has been reported by many groups.\cite{jun,moto,bayr,balsys} Magnetic ordering has been observed for Na$_{x}$CoO$_{2}$ with low transition temperature of $\sim$ 22 K (T$_{m}$).\cite{jun,moto,booth,dj,dj2,foo} The study of Sugiyama \textit{et al.}\cite{jun} suggests that the state of Na$_{0.75}$CoO$_{2}$ is either a ferrimagnet or a commensurate spin-density-wave below T$_{m}$ (22 K). On the other hand, Motohashi \textit{et al.}\cite{moto} study suggests a possible charge-density-wave or a spin-density-wave state in this compound. Moreover, the existence of ferromagnetic spin fluctuations within the cobalt-oxygen layer have been reported by Boothroyd \textit{et al.}.\cite{booth} D. J. Singh\cite{dj,dj2} has predicted the ferromagnetic instability in Na$_{x}$CoO$_{2}$ (x = 0.3 to 0.7) system using density functional theory within local spin density approximation (LSDA). Foo \textit{et al.}\cite{foo} reported the ground state of Na$_{x}$CoO$_{2}$ (x = 0.65 – 0.75) as Curie-Weiss metallic. Therefore, from the above discussion it is clear that the low temperature magnetic phase is controversial in this system. Here, we want to mention that we have studied the transport properties of Na$_{0.74}$CoO$_{2}$ above 300 K, where the compound possesses paramagnetic phase. Thus, the exact knowledge about the low temperature magnetic phase (which turns out to be controversial) is not important for our purpose, if the calculated electronic structure provides the reasonably accurate electronic structure corresponding to paramagnetic phase.


Computational approach alwayes help in either searching the new materials or explaining the existing experimental results. In computational methods, density functional theory (DFT)\cite{kohn1,kohn2} is one of the most popular tool to predict the ground state electronic structure of the materials. In DFT, many exchange-correlation (XC) functionals have been developed for better approximations with their own merit and demerit.\cite{shastri} However, DFT is not capable of explaining the electronic structure of strongly correlated electron systems containing partially filled \textit{d} or \textit{f}-band. In this context, DFT+\textit{U} can be used, where on-site Coulomb interaction,\textit{U} takes the responsibility to overcome this problem. Therefore, the motivation of this work is up to what extent the experimentally observed TE properties of Na$_{0.74}$CoO$_{2}$ can be explained by DFT and DFT+\textit{U}. Then by taking the best computational results, can we further improve the TE properties of the compound in high temperature region where performing our experiment is not accessible?      

In this work, we have studied the thermoelectric properties of Na$_{0.74}$CoO$_{2}$ using experimental and theoretical approaches. Spin-polarized DFT + \textit{U} (= 4 eV) calculation is found to give the best matching with the temperature dependent experimental data. The maximum calculated ZT value is found to be $\sim$ 0.67 for p-type conduction, whereas it reaches $\sim$ 2.7 for n-type conduction. The maximum possible value of efficiency ($\eta$) of TEG (which is made by p and n-type of this compound) is found to be $\sim$ 11$\%$ when the cold and hot end temperatures were fixed to be at 300 K and 1200 K, respectively. These high temperature values of ZT and efficiency suggest that this compound can be used as a high temperature TE applications in power plant utility.             

\section{Experimental and computational details}

Na$_{0.74}$CoO$_{2}$ sample was synthesized by solid state reaction, where Na$_{2}$CO$_{3}$ and Co$_{3}$O$_{4}$ were used as starting materials. Then, the powder were sintered directly at 1073 K for 12 hours in a preheated furnace. Finally, for doing the measurement, we have made the pellet and sintered at 1073 K for 24 hours. More details of synthesis procedure and characterization can be found in Ref. [34]. The measurements of S and $\kappa$ were performed in the temperature range $300-620$ K by using the home-made experimental setup\cite{S,k,tot}. The circular pellet of 8 ($\pm$0.1) mm diameter with 1 ($\pm$0.02) mm thickness was used for S and $\kappa$ measurements.

To examine the experimentally observed transport properties of the sample, we have done electronic structure calculations with DFT.\cite{kohn1,kohn2} In the calculations, full-potential linearized augmented plane wave (FP-LAPW) method is used which is implemented in WIEN2k code.\cite{wein2k} The PBEsol\cite{pbesol} is used as an exchange correlation functional within generalized gradient approximation (GGA). The experimental lattice parameters of a = b = 2.823 \AA\, \& c = 10.933 \AA\, and space group: P63/mmc (no. 194) were used for the ground-state self-consistent calculations. The muffin-tin sphere radii for Co, Na and O atoms were fixed to be 1.79, 2.09 and 1.54 Bohr, respectively. The on-site coulomb interaction srength, U was taken from 2 to 5 eV in DFT+\textit{U} method. A k-mesh of size $3\times15\times8$ was used for the calculations. The criteria for calculating the total ground state energy of the compound in self-consistency cycle was set to be less than 0.1 mRy. For such a vacant compound minimization of internal parameters have done to relax the system. Force convergence criteria was set to be 1 mRy/Bohr during the minimization of internal  parameters. The temperature dependence of transport properties (S and $\sigma$) are calculated by using BoltzTraP package\cite{boltztrap}, which is interfaced with WIEN2k.\cite{wein2k} A k-mesh of size $19\times75\times43$ was used for calculating the transport properties of the compound. It is important to note that this integration mesh appears to be strange. Since, the calculation has been carried out on a supercell of $2\times1\times1$ and the lattice parameters of final structure are found to be 20.66, 5.34, 9.24 Bohr. For this structure we have used 64000 k-points in the full Brillouin zone and accordingly for even sampling of the k-points along the x, y and z directions in the irreducible part of the Brillouin zone, program automatically finds the k-mesh of $19\times75\times43$.

\section{Results and Discussion}

\subsection{\label{sec:level2}Experimental transport properties}
Fig. 1(a) shows the experimentally observed Seebeck coefficients (S) of Na$_{0.74}$CoO$_{2}$ compound as a function of temperature. The measured value of S is $\sim$ 64 $\mu$V/K at 300 K which is found to give a good agreement with reported values for Na$_{0.7}$CoO$_{2}$.\cite{kaurav,altin} From Fig. 1(a), it is clear that as the temperature increases, the values of S also increase almost linearly up to the highest temperature studied. At 620 K, the value of S is found to be $\sim$ 118 $\mu$V/K. The positive values of S are observed in the full temperature range indicating the dominating p-type behaviour of the compound. The value of S at 620 K is almost double the value at room temperature. In this context, it should be mentioned that many thermoelectric materials show non-monotonic temperature dependence of S where the magnitude of S first increases and then decreases with rise in temperature in some temperature range.  For example, in Bi$_{2}$Te$_{3}$, the value of S increases up to $\sim$ 400 K and then decreases.\cite{yang} In our case, the increasing nature of S with temperature suggesting that this material can give rise to high power factor (S$^{2}\sigma$) at high temperature region and our calculations also support it which is discussed later. 
	
\begin{figure} 
\includegraphics[width=0.90\linewidth, height=13.0cm]{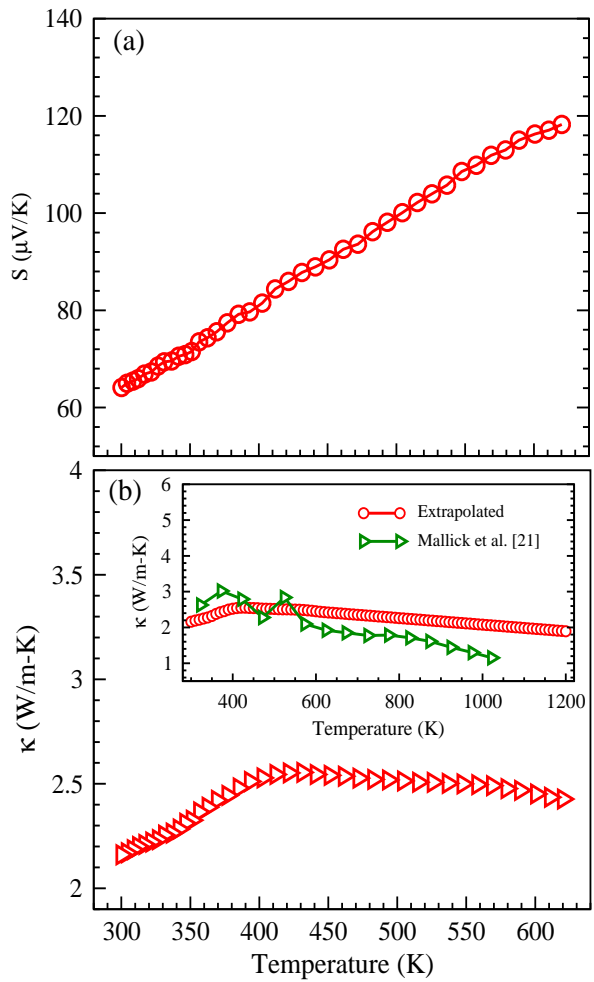} 
\caption{\small{Temperature dependence of (a) Seebeck coefficient and (b) Thermal conductivity of Na$_{0.74}$CoO$_{2}$ compound.}}
\end{figure}     

Temperature dependence of thermal conductivity ($\kappa$) of the sample in the temperature region $300-620$ K is displayed in Fig. 1(b). The observed value of $\kappa$ is $\sim$ 2.2 W/m-K at 300 K, which found to give a good match with reported value by Mallick \textit{et al.}\cite{vitta}. In the temperature range $300-430$ K, the value of $\kappa$ increases up to $\sim$ 2.6 W/m-K at 430 K. As the temperature increases further from 430 K, the value of $\kappa$ is decreasing slowly and it's value is found to be $\sim$ 2.4 W/m-K at 620 K. These low values of $\kappa$ under the studied temperature range help to increase the ZT value for this compound. Here, we have extrapolated the $\kappa$ values up to 1200 K for estimating the high temperature ZT values. These extrapolated κ values are also compared with reported data by Mallick \textit{et al.}\cite{vitta} as shown in inset of Fig. 1 (b).
At lower temperature one can notice fairy good agreement in both the data. However, at higher temperature both the data started deviating where this deviation keeps on increasing with increase in temperature. 

\subsection{\label{sec:level2}Force minimization and electronic structure calculations} 
The primitive cell of Na$_{0.74}$CoO$_{2}$ compound contains 10 atoms to form a lattice with the Wyckoff positions: Co at 2a(0, 0, 1/2), Na1 at 2b(0, 0, 1/4), Na2 at 2c(2/3, 1/3, 3/4) and O at 4f(1/3, 2/3, 0.0913). The formula unit contains one Co atom, 0.23 Na1 atom, 0.51 Na2 atom and two O atoms. Therefore, the primitive cell contains two formula unit with Na vacancy in both the sites (2b and 2c). For the sake of simplicity, we have considered 0.25 Na1 atom and 0.50 Na2 atom in the formula unit which is corresponding to Na$_{0.75}$CoO$_{2}$. For doing the calculation, we have created the supercell of $2\times1\times1$ which contains 4 Na1 atoms as well as 4 Na2 atoms. Then, 3 Na1 atoms and 2 Na2 atoms have been removed from  the supercell to get this compound. This provides various possibilities for removing the Na atoms from both the sites. We tried possible 24 combinations which give almost same results once we relax the structure by giving the force minimization criteria in self-consistent calculations. Therefore, care should be taken to do the electronic structure calculations for such a vacant system. Kurniawan \textit{et al.}\cite{kurniawan} have performed  the DFT calculations on Na$_{x}$CoO$_{2}$ (x = 0.2 to 1) by creating the supercell of $2\times2\times1$ to simulate the variation of Na content using PBE.

The electronic structure calculations were performed by using spin-unpolarized (SUP) DFT and spin-polarized (SP) DFT + \textit{U} (= 4 eV) methods. The selected bond distances and angles are shown in Table I corresponding to SUP and SP results along with experiment. Fig. 2 shows the crystal structure of the compound of 2$\times$1$\times$1 supercell with indexing the atoms for supporting of Table I. From the table it is seen that all the $Co-O$ bond lengths corresponding to SUP and SP results are different from that of experiment which may be due to supercell calculations. The average distance of $Co-O$ for SUP and SP are 1.903 \AA\, and 1.904 \AA, respectively and which are very much close to experimental value of 1.912 \AA. The average bond angle of $Co-O-Co$ corresponding to SUP and SP results are 95.51 deg and 95.49 deg, respectively and which are also closely matching with experimental value of 95.22 deg. The value of total energy in the case of SP calculation is found to be $\sim$ 7.5 meV/f.u. lower than that of SUP calculation. This result suggests that the SP calculation gives the ground state of this compound. The calculated magnetic moment of this compound is also found to be $\sim$ 1.12 $\mu_{B}$/f.u., which gives the good agreement with the experimentally observed values of magnetic moment.\cite{altin,viciu} The major contribution in magnetic moment comes from Co atom which is almost 80$\%$ of the total value. In order to explain the experimentally observed transport properties of this compound, the density of states (DOS) calculations are carried out. 

\begin{table}
\caption{\small{Selected bond lengths and angles corresponding to spin-unpolarized (SUP) and spin-polarized (SP) phases of Na$_{0.74}$CoO$_{2}$ along with experiment.}}
\resizebox{0.45\textwidth}{!}{%
\begin{tabular}{@{\extracolsep{\fill}}c c c c c c c c c c c} 
\hline\hline
 
\multicolumn{1}{c}{Type of bond} & & & & &  \multicolumn{4}{c}{Bond length (\AA)}  &  \multicolumn{1}{c}{} & \multicolumn{1}{c}{}\\ 
\cline{4-11}                                
  & & & \multicolumn{1}{c}{SUP} & & & \multicolumn{1}{c}{SP}   &   \multicolumn{1}{c}{Experiment}\\
   
 \hline
$Co(1)-Co(2)$ & & & 2.8232 & & &  2.8231   & 2.8237 \\
$Co(1)-O(1)$ & & & 1.9297 & & &  1.9294   & 1.9116 \\
$Co(1)-O(2)$ & & & 1.8810 & & &  1.8822   & 1.9116 \\
$Co(1)-O(3)$ & & & 1.9256 & & &  1.9261   & 1.9116 \\
$Co(1)-O(4)$ & & & 1.8658 & & &  1.8655   & 1.9116 \\
$Co(2)-O(1)$ & & & 1.9313 & & &  1.9306   & 1.9116 \\
$Co(2)-O(2)$ & & & 1.8868 & & &  1.8875   & 1.9116 \\
\hline\hline
\multicolumn{1}{c}{Type of bond} & & & &  \multicolumn{4}{c}{Bond angle (deg)} & \multicolumn{1}{c}{} & \multicolumn{1}{c}{}\\ 
\cline{4-11}                                
  & & & \multicolumn{1}{c}{SUP} & & & \multicolumn{1}{c}{SP}   &   \multicolumn{1}{c}{Experiment}\\
\hline
$Co(1)-O(1)-Co(2)$ & & & 93.975 & & & 94.001   & 95.220 \\
$Co(1)-O(2)-Co(2)$ & & & 97.054 & & & 96.987   & 95.220 \\
\hline\hline
 
\end{tabular}}
\end{table}  

\begin{figure} 
\includegraphics[width=0.65\linewidth, height=6.3cm]{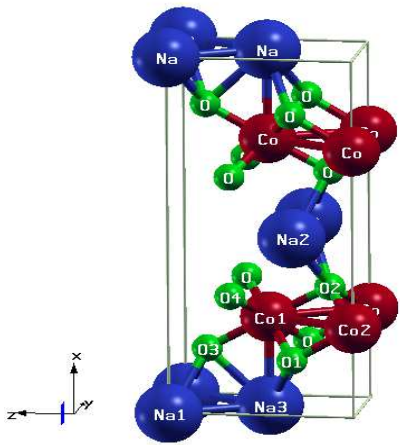} 
\caption{\small{Crystal structure of Na$_{0.74}$CoO$_{2}$ for 2$\times$1$\times$1 supercell.}}
\end{figure}

\begin{figure*} 
\includegraphics[width=0.90\linewidth, height=7.6cm]{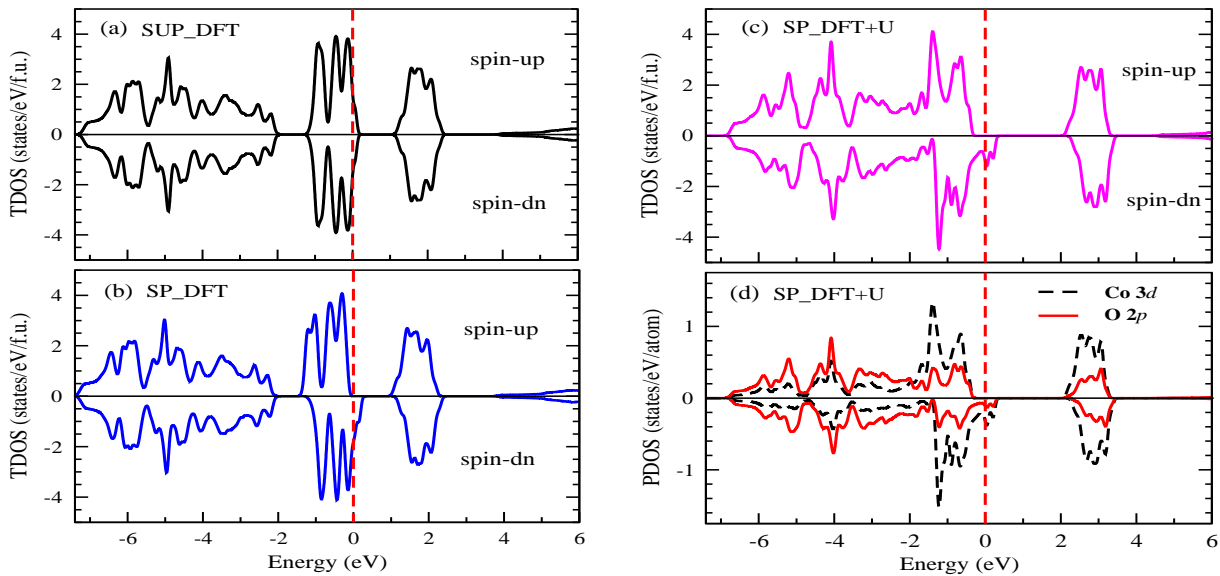} 
\caption{\small{Total density of states (TDOS) of (a) spin-unpolarized (SUP) solution using DFT, (b) spin-polarized (SP) solution using DFT and (c) SP solution using DFT+$U$; (d) Partial DOS (PDOS) of SP solution using DFT+$U$.}}
\end{figure*}

Fig. 3 exhibits the DOS plots corresponding to SUP and SP calculations of Na$_{0.74}$CoO$_{2}$. The dashed line at 0 eV represents the Fermi level, E$_{F}$ of the compound. Fig. 3(a) shows the total DOS (TDOS) for SUP solution within DFT. The value of DOS is found to be $\sim$ 3.2 states/eV/f.u. at Fermi level. This considerable amount of DOS at E$_{F}$ may be considered as a signature of a magnetic ground state of the compound on the basis of Stoner theory. TDOS corresponding to SP solution is also calculated under DFT, which is shown in Fig. 3(b). The contribution of DOS corresponding to spin-up and spin-dn channels are asymmetric. From the figure it is observed that spin-dn channel is giving finite DOS of $\sim$ 2 states/eV/f.u. at E$_{F}$, whereas spin-up channel gives the finite band gap of $\sim$ 0.9 eV. Therefore, this result predicts the half-metallic ground state of the compound which is in consistent with the results of earlier work carried out in Na$_{x}$CoO$_{2}$ system.\cite{dj,casolo}

In order to explain the experimental results in better way, inclusion of on-site Coulomb interaction for Co 3\textit{d} electrons in DFT+\textit{U} calculations is required. The TDOS has calculated for different values of \textit{U} varied from 2 to 5 eV. Then the value of \textit{U} is chosen to be 4 eV by comparing the calculated transport results with experiment. This value of \textit{U} is close to the value of \textit{U} used by Wissgott \textit{et al.}\cite{wissgott} in their study of local-density approximation (LDA) + dynamical mean-field theory (DMFT) for Na$_{0.74}$CoO$_{2}$. Fig. 3(c) shows the calculated TDOS plot of the compound corresponding to SP solution under DFT+\textit{U} method. Inclusion of \textit{U}, generally increase the separation between occupied and unoccupied states and transfers the spectral weight between the energy positions. Due to which, gap in the energy region $\sim$ -1.85 to -1.33 eV vanishes. From the figure, it is also seen that the features of DOS in dn-channel around E$_{F}$ is different from that of DFT DOS. Therefore, all these results suggest that DFT+\textit{U} calculation is expected to give the different transport properties over DFT for this compound.  

In order to know the contribution in TDOS from different atomic orbitals, we have calculated the partial DOS (PDOS) for constituent atoms of Na$_{0.74}$CoO$_{2}$. From the calculations, it is observed that the electronic contributions near the Fermi level comes from Co 3\textit{d} and O 2\textit{p} orbitals mainly with negligibly small contribution from Na 3\textit{s} orbital. Hence, here we present the PDOS corresponding to Co 3\textit{d} and O 2\textit{p} orbitals, which is shown in fig. 3(d). From the figure, it is clear that at E$_{F}$, a quite good amount of DOS in dn-channel is observed. These values are found to be $\sim$ 1.31 states/eV/atom and $\sim$ 0.45 states/eV/atom for Co 3\textit{d} and O 2\textit{p} orbitals, respectively. Thus, due to metallic nature of dn-channel, the electrons from hybridized Co 3\textit{d} and O 2\textit{p} states are expected to give the large contribution in electrical conductivity ($\sigma$). In specific, the contributions in DOS from Co 3\textit{d} orbitals is almost triple than that of O 2\textit{p} orbitals in spin-dn channel at E$_{F}$. This suggests that Co 3\textit{d} orbitals take the main responsibility in $\sigma$ values. For up-channel, from the edge of the valence band (VB), peaks are associated at $\sim$ $-$ 0.65 eV for both Co 3\textit{d} and O 2\textit{p} orbitals and the values of DOS are found to be $\sim$ 3.58 states/eV/atom and $\sim$ 1.75 states/eV/atom, respectively. Therefore, due to semiconducting nature of up-channel, large values of S and small values of $\sigma$ are expected from Co 3\textit{d} and O 2\textit{p} orbitals. Here, one should conclude that the major contribution in S and $\sigma$ come from Co 3\textit{d} mainly.

\begin{table*}
\caption{\label{tab:table1}%
\small{Calculated Seebeck coefficients (S) at different temperature by using DFT and DFT+$U$. The average of absolute deviation of S from experiment is also shown, where n is the total number of data points at different temperature. (Here, S is in $\mu$V/K unit).
}}
\begin{ruledtabular}
\begin{tabular}{lcccccc}
\textrm{Methods}&
\textrm{S at 300 K}&
\textrm{S at 400 K}&
\textrm{S at 500 K}&
\textrm{S at 600 K}&
\textrm{S at 620 K}&
\textrm{$(\sum|S_{exp}-S_{cal}|)/n$}\\
      
\colrule
Exeriment & 64.08 & 80.98 & 99.30 & 116.18 & 118.21 & --- \\
DFT       & 58.12 & 81.47 & 98.80 & 111.43 & 113.54 & 2.03 \\
DFT+$U$(= 2 eV)    & 59.05 & 81.35 & 98.60 & 111.00 & 113.01 & 2.06 \\
DFT+$U$(= 3 eV)    & 57.88 & 79.40 & 97.05 & 110.21 & 112.39 & 3.45 \\
DFT+$U$(= 4 eV)    & 64.43 & 82.75 & 99.55 & 112.96 & 115.23 & 1.41 \\
DFT+$U$(= 5 eV)    & 59.16 & 78.68 & 96.35 & 109.90 & 112.11 & 3.82 \\
\end{tabular}
\end{ruledtabular}
\end{table*}

\begin{figure} 
\includegraphics[width=0.95\linewidth, height=7.5cm]{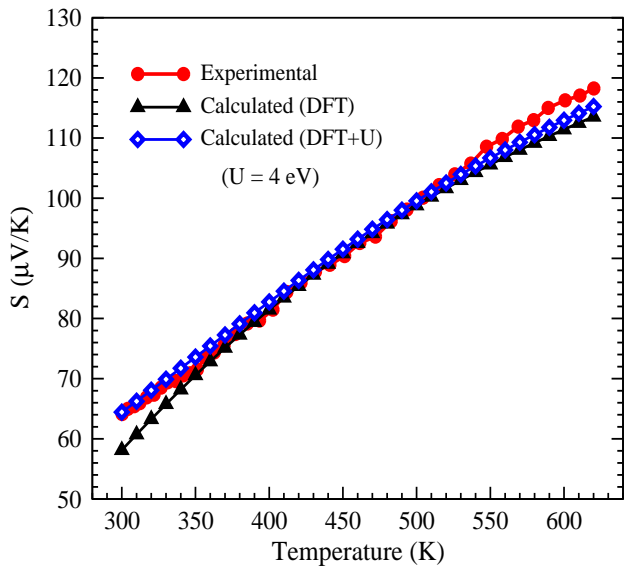} 
\caption{\small{Comparison of experimental and calculated (using DFT and DFT+\textit{U}) values of Seebeck coefficient (S) as a function of temperature.}}
\end{figure}

\subsection{\label{sec:level2}Calculated transport properties} 

The temperature dependence of S and $\sigma$ are calculated for spin-up and spin-dn channels using BoltzTraP package\cite{boltztrap}. The total S and $\sigma$ values are needed to be calculated from up and dn-channels to analyze the experimental results. Here, two current model is employed to calculate the total S from the values of spin-up and spin-dn channel. In this model, total S can be expressed as the following equation\cite{current1, current2, sharma, saurabh}
\begin{equation}
S = \frac{(\sigma^{\uparrow}S^{\uparrow}+\sigma^{\downarrow}S^{\downarrow})}{(\sigma^{\uparrow}+\sigma^{\downarrow})}
\end{equation}
where, $\sigma^{\uparrow}$ and $\sigma^{\downarrow}$ are the electrical conductivities and S$^{\uparrow}$ and S$^{\downarrow}$ are the Seebeck coefficients of up and dn-channels, respectively. Using Eqn 2, the values of S are calculated within DFT and DFT+\textit{U} methods. At $\mu$ = 0 eV (Fermi level), the calculated values of S are  $\sim$ $-$15 to 19 $\mu$V/K at 300 K using DFT and DFT+\textit{U} and which are quite far away from the experimental value of $\sim$ 64 $\mu$V/K. This result suggests that the $\mu$ value may not be suitable for calculating S, which can be understood by the following ways. Here, we have performed the calculation on stoichiometric Na$_{0.75}$CoO$_{2}$ compound. However, the real sample on which the experiment has been carried out may have excess/vacant oxygen along with Na vacancy. Presence of oxygen non-stoichiometry may lead to change in the chemical potential. The quantification of the exact value of the $\mu$ in this situation is not straight forward. We simply tried to find out a single value of $\mu$ at 300 K which provides best representation of the temperature dependent experimental data. We found that at $\mu$ = 143, 182, 198, 220 and 231 meV, the calculated values of S are $\sim$ 58 (DFT), $\sim$ 59 (\textit{U} = 2 eV), $\sim$ 58 (\textit{U} = 3 eV), $\sim$ 64 (\textit{U} = 4 eV) and $\sim$ 59 (\textit{U} = 5 eV) $\mu$V/K, respectively at 300 K and give the best match in each case. Using these $\mu$ values, we have calculated S in the temperature region $300-620$ K as shown in Table 2. The average of absolute deviation of calculated S from experiment are also shown in the table. It is seen from the table that the lowest deviation is observed when S was calculated using \textit{U} = 4 eV. Fig. 4 shows the calculated values of S using DFT and DFT +\textit{U} (= 4 eV) along with experiment. Our calculated value of S at 300 K using DFT + \textit{U} (= 4 eV) is close to the value of 66 $\mu$V/K at 290 K calculated by Wissgott \textit{et al.}\cite{wissgott} using LDA + DMFT. From Fig. 4, it is evident that the calculated S values using DFT is deviating below $\sim$ 360 K and above $\sim$ 500 K from the experimental data which may be due to DFT is not able to give a proper DOS around the Fermi level for this compound as we have seen in DOS plots. From the figure it is clearly seen that the calculated values of S (using \textit{U} = 4 eV) are giving good match with experiment throughout the studied temperature range over DFT. This value of \textit{U} is in accordance with the reported value by Wissgott \textit{et al}.\cite{wissgott}. The change in S with $\mu$ is calculated at different temperature using \textit{U} = 4 eV, as shown in Fig. 5 and which is used in finding out the $\mu$ value for calculating the temperature dependence of S. From the figure (left inset) it is clear that the positive values of S are seen at $\mu$ $\approx$ 220 meV for all the studied temperatures which is indicating the p-type behaviour of the compound and which is in accordance with our experimental results. Therefore, \textit{U} = 4 eV can be used to study the temperature dependence of transport properties of Na$_{0.74}$CoO$_{2}$. For rest of the calculations, we used \textit{U} = 4 eV and $\mu$ = 220 meV in this work. 

\begin{figure} 
\includegraphics[width=1.0\linewidth, height=7.5cm]{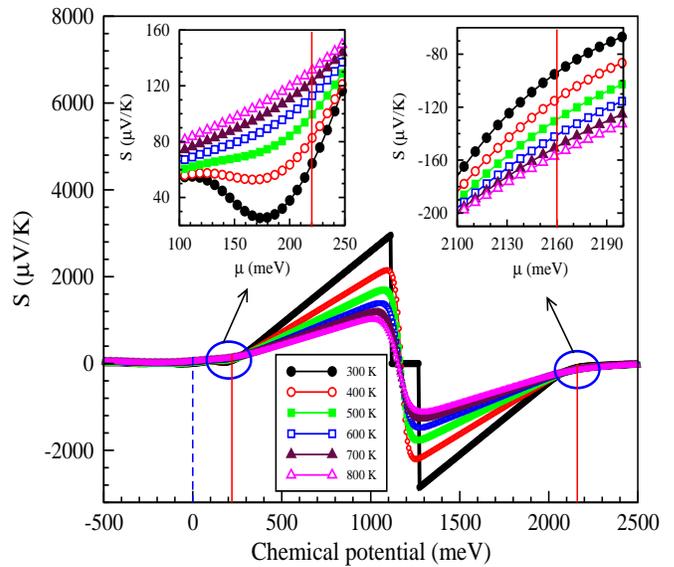} 
\caption{\small{Change in Seebeck coefficient (S) with chemical potential at different temperature using DFT+\textit{U} (= 4 eV).}}
\end{figure}     

\begin{figure} 
\includegraphics[width=0.95\linewidth, height=7.5cm]{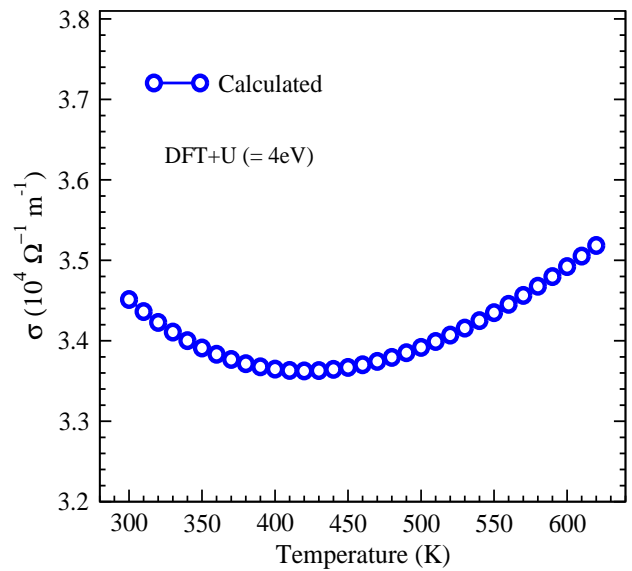} 
\caption{\small{Calculated values of electrical conductivity ($\sigma$) as a function of temperature using DFT+\textit{U} (= 4 eV).}}
\end{figure}

\begin{figure*} 
\includegraphics[width=0.88\linewidth, height=7.0cm]{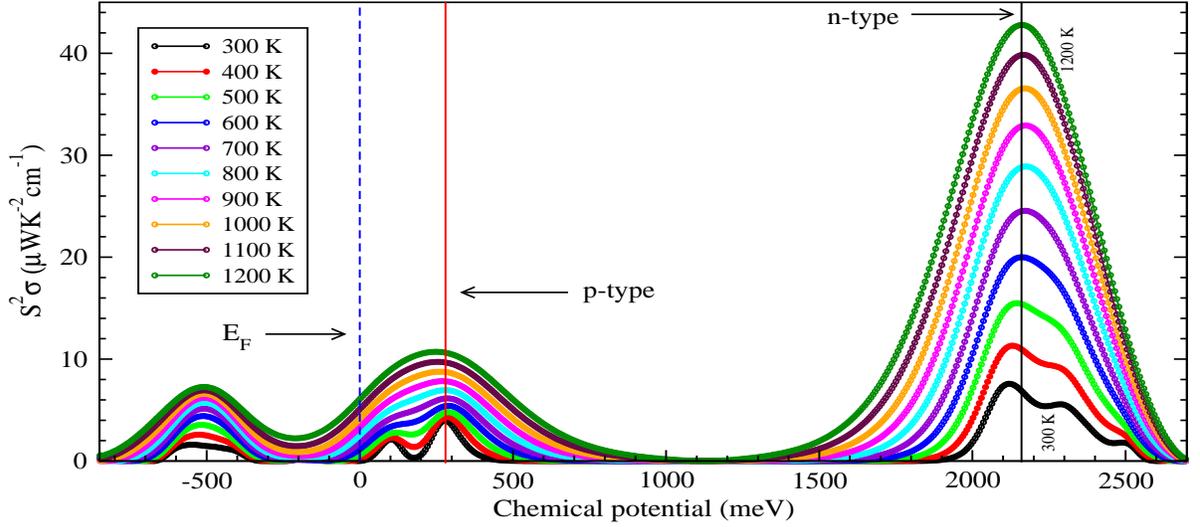} 
\caption{\small{Variation of power factor (S$^{2}\sigma$) with chemical potential at different temperatures.}}
\end{figure*}

\begin{figure*} 
\includegraphics[width=0.93\linewidth, height=7.2cm]{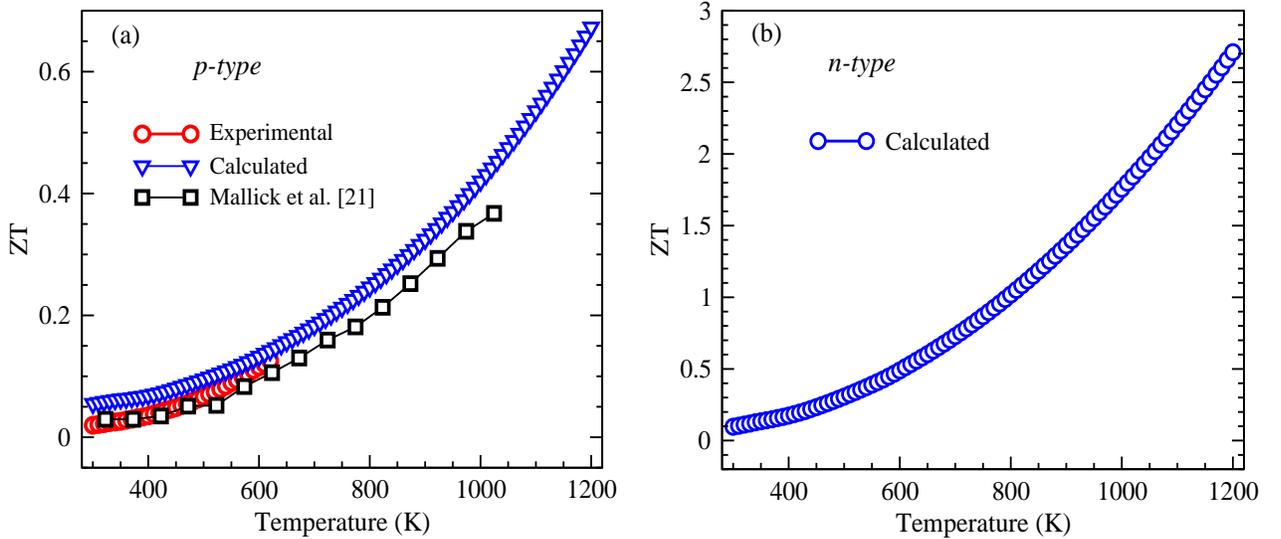} 
\caption{\small{Temperature dependence of \textit{figure-of-merit} for (a) p-type and (b) n-type Na$_{0.74}$CoO$_{2}$.}}
\end{figure*} 

The temperature dependence of $\sigma$ is also calculated for up and dn-channels. According to two-current model, the total $\sigma$ can be calculated by the sum of $\sigma$ of spin-up and spin-dn channels and expressed as\cite{current3}
\begin{equation}
\sigma = \sigma^{\uparrow}+\sigma^{\downarrow}
\end{equation}
The values of $\sigma$ are calculated using Eqn. 3. But, we got $\sigma^{\uparrow}$ ( and $\sigma^{\downarrow}$) per relaxation time ($\tau$) from calculation. Therefore, the value of $\tau$ is needed to be calculated to get an actual $\sigma$. Comparing the reported values of $\sigma$ (3.45 $\times$ 10$^{4}$ $\Omega^{-1}$ m$^{-1}$) at 300 K by Altin \textit{et al}\cite{altin} with calculated $\sigma/\tau$, we have calculated the $\tau$ value as $0.7\times10^{-14}$ s. This calculated value of $\tau$ at RT is in the typical range of $10^{-14}-10^{-15}$ for metals and semiconductors.\cite{ashcroft} Then this constant value of $\tau$ is used to calculate the $\sigma$ in the temperature region $300-620$ K as shown in Fig. 6 and also used for rest of the calculations in the present work. From the figure it is seen that the values of $\sigma$ decrease up to $\sim$ 420 K. After 420 K, these values are increasing up to 620 K with the corresponding value of 3.52 $\times$ 10$^{4}$ $\Omega^{-1}$ m$^{-1}$.   

\begin{figure*} 
\includegraphics[width=0.95\linewidth, height=7.0cm]{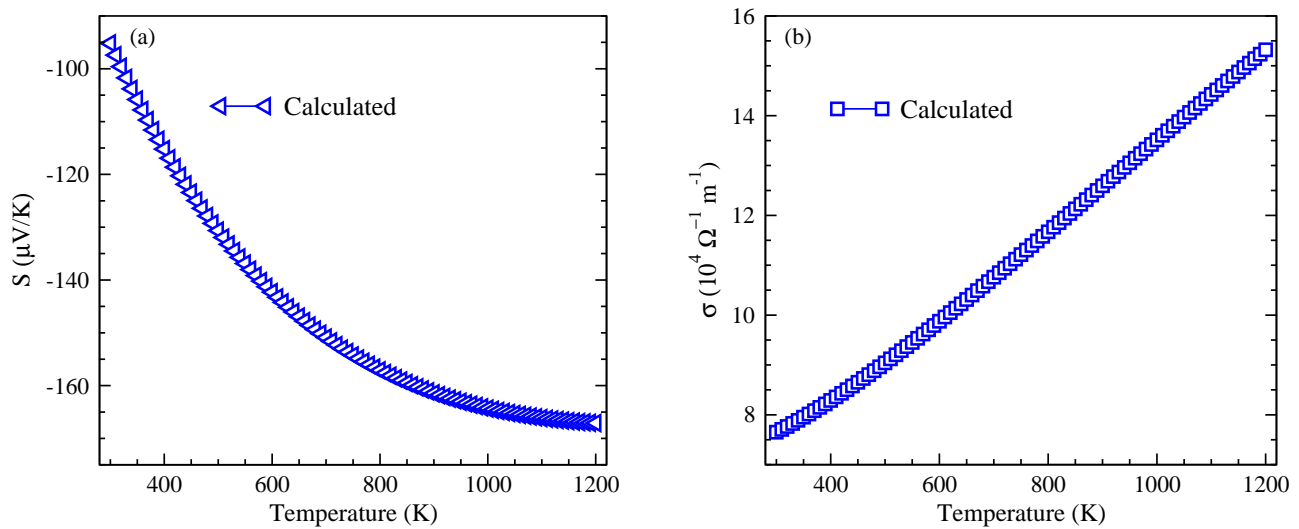} 
\caption{\small{Calculated values of (a) Seebeck coefficients and (b) Electrical conductivity of n-type Na$_{0.74}$CoO$_{2}$} in the temperature region $300-1200$ K using DFT +\textit{U} (= 4 eV).}
\end{figure*}

\subsection{\label{sec:level2}Power factor, \textit{figure-of-merit} and efficiency}

To see the applicability of Na$_{0.74}$CoO$_{2}$ in TE world, we have calculated the power factor (PF) with $\mu$ at different temperature which is shown in Fig. 7. The dotted line at 0 eV denotes the Fermi level, E$_{F}$ of the compound. Around the E$_{F}$, the maximum PF is observed at $\sim$ 280 meV for p-type behaviour with the corresponding values of $\sim$ 4.2 $\mu$WK$^{-2}$cm$^{-1}$ and $\sim$ 10.5 $\mu$WK$^{-2}$cm$^{-1}$ for 300 K and 1200 K, respectively. Here, it is important to note that experimental S and $\sigma$ was matched with calculated values at $\mu$ $\approx$ 220 meV. Therefore, this calculation of PF suggest that small amount of doping may gain the capability in TE application of Na$_{0.74}$CoO$_{2}$.


By using the experimentally measured S and $\kappa$ values with calculated $\sigma$, we have computed the \textit{figure-of-merit} (ZT) using Eqn. 1 as given in Fig. 8(a) in the temperature region $300-620$ K. The values of ZT increase with increasing of temperature. The observed value of ZT at 300 K is found to be $\sim$ 0.02, whereas it reaches $\sim$ 0.13 at 620 K. These values of ZT give good agreement with the reported values by Mallick \textit{et al.}\cite{vitta} The ZT value at 620 K is $\sim$ 7 times larger than that of room temperature value. This result suggests to give an attention in high temperature TE application of this compound. Here, it is important to note that there is a scope to further improve the value of ZT by doping to achieve $\mu$ $\approx$ 280 meV, where PF is maximum (Fig. 7). By taking these values of PF along with experimental $\kappa$ values, we have calculated ZT values in the temperature region $300-1200$ K which is also presented in Fig. 8(a). The calculated value of ZT at 300 K is $\sim$ 0.05, whereas it is found to be $\sim$ 0.67 at 1200 K. In the full temperature region, ZT values are increasing monotonically, which are compared with the reported data by Mallick \textit{et al.}\cite{vitta} This comparison gives good match throughout their studied temperature range ($\sim$ $320-1025$ K) as shown in Fig 8(a). The values of ZT at high temperature are large enough to be used in high temperature TE applications. 

Till now, our entire discussion was on TE properties of p-type Na$_{0.74}$CoO$_{2}$. As we know that for making the thermoelectric generator (TEG), we need both p-type and n-type materials. Therefore, for the sake of curiosity, we examined the DOS plot to see the possibility for making this compound as n-type. If we observe the DOS plot of Fig. 3(c), we can see that by small amount of doping one can reach in semiconductor region (which starts at $\sim$ 385 meV) as we are already at 220 meV above the E$_{F}$. Now, by doping of electrons one can make this compound as n-type as shown in Fig. 5 (right inset). Temperature dependence of PF are calculated in n-type region to find out the maximum value as shown in Fig. 7. From the figure, it is found that the maximum PF can be obtained at $\sim$ 2160 meV, which corresponds to the electron doping of $\sim$ 5.1$\times$10$^{20}$ cm$^{-3}$. At this doping level, the temperature dependence of S and $\sigma$ are calculated in temperature region $300-1200$ K as shown in Fig. 9(a) and 9(b), respectively. The negative values of S in the full temperature region indicate the dominating n-type behaviour of the compound. From the figure it is clear that the magnitude of S as well as the value $\sigma$ are high as compared to p-doped compound. This may be due to the (i) presence of more charge carriers leading to increase in $\sigma$ and (ii) the band structure around the chemical potential 2160 meV responsible for increasing the value of S. These two effects  drive the power factor (S$^{2}\sigma$) so high for the n-doped case. By taking these values of S and $\sigma$ along with extrapolated experimental $\kappa$, we have estimated the ZT values in the temperuture region $300-1200$ K which are presented in Fig. 8(b). The values of ZT at 300 K and 1200 K are found to be $\sim$ 0.1 and $\sim$ 2.7, respectively. At this stage, it is important to note that the values of ZT are calculated using the extrapolated experimental $\kappa$ values measured for p-type Na$_{0.74}$CoO$_{2}$. Here, it should be mentioned that the $\tau$ value calculated at 300 K was used for calculating the temperature dependence of $\sigma$. But, at high temperature, in general, $\tau$ has temperature dependence due to electron-phonon interactions.\cite{ahmad} Moreover, $\tau$ has also energy dependence which may make it $\mu$ dependence. Thus, one should keep in mind that the actual value of electrical conductivity is expected to be lower than the calculated value in the high temperature region. Due to this, one may get ZT value less than 2.7 at 1200 K, if this compound is synthesized with suitable n-type doping as mentioned. Therefore, high values of ZT of this compound in high temperature region deserved attention of the community in TE world.  

Here, we have explored the possibilities of making n-doped compounds. To look at these possibilities, we have performed the calculations for doping of Ni at Co site and creating oxygen vacancy. Fig. 10 shows the calculated DOS for 6.25$\%$ of Ni doping at Co site and 3.125$\%$ of oxygen vacancy. One can see the presence of finite impurity states in the gaped region above $\sim$ 0.5 eV. This suggests that one can get the predicted thermoelectric properties either by doping Ni at Co  sites or by creating oxygen vacancies in appropriate amount.  Doping of Ni is expected not to change the structural stability as the ionic radius of Co$^{3+}$ and Ni$^{3+}$ are almost same ($\sim$ 55 pm). Oxygen vacancy of $\sim$ 3$\%$ is also not expected to disturb the structural stability significantly.

Finally, we have calculated the maximum possible efficiency of TEG which is supposed to be made by p and n-type Na$_{0.74}$CoO$_{2}$ as shown in Fig. 11. Here, the cold end temperature (T$_{c}$) was kept constant at 300 K, whereas the hot end temperatures (T$_{h}$) were varied from 300 to 1200 K. The segmentation method was employed to calculate the efficiency as implemented by Gaurav \textit{et al.}\cite{gaurav}. In this method, the efficiency of every segment is calculated by the following expression,\cite{sherman} 

\begin{figure} 
\includegraphics[width=0.94\linewidth, height=7.1cm]{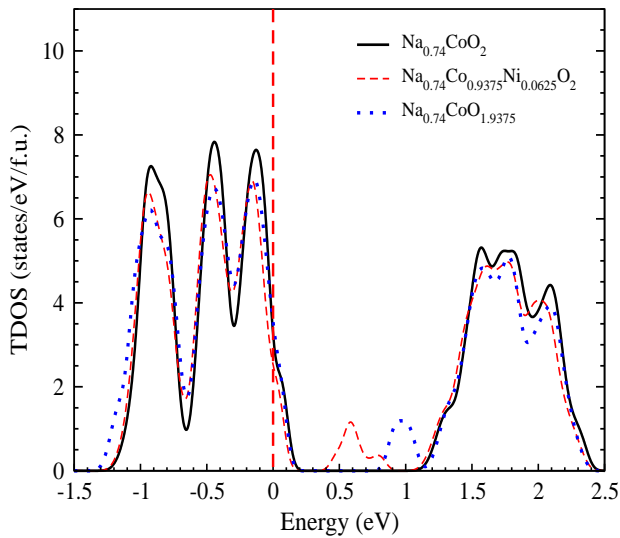} 
\caption{\small{Efficiency ($\eta$) of p and n-type Na$_{0.74}$CoO$_{2}$ with hot end temperature, keeping cold end fixed at 300 K.}}
\end{figure}

\begin{figure} 
\includegraphics[width=0.9\linewidth, height=7.0cm]{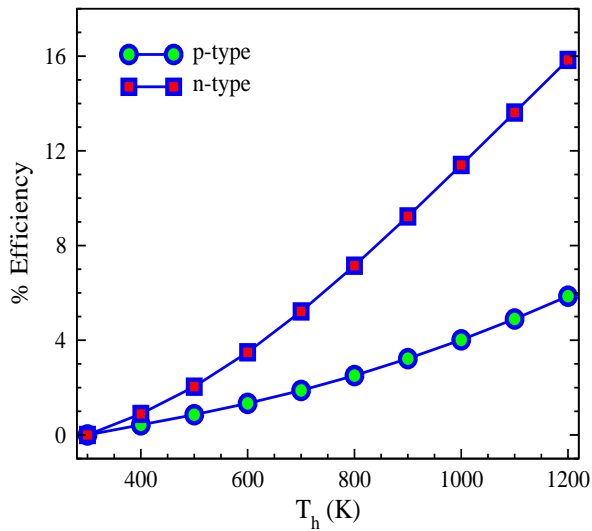} 
\caption{\small{Total density of states (TDOS) of Na$_{0.74}$CoO$_{2}$, Na$_{0.74}$Co$_{0.9375}$Ni$_{0.0625}$O$_{2}$ and Na$_{0.74}$CoO$_{1.9375}$.}}
\end{figure}

\begin{equation}
\eta_{i} = \dfrac{\Delta T}{\overline{\rm T}+\dfrac{\Delta T}{2}}\dfrac{\sqrt{1+Z\overline{\rm T}}-1}{\sqrt{1+Z\overline{\rm T}}+\dfrac{\overline{\rm T}-\dfrac{\Delta T}{2}}{\overline{\rm T}+\dfrac{\Delta T}{2}}}
\end{equation}
where $\Delta T$ and $\overline{\rm T}$ is the temperature difference and average temperature of each individual segment, respectively. First of all, we will discuss the efficiency of p-type Na$_{0.74}$CoO$_{2}$ and then n-type. In the temperature region $300-1200$ K, the total number of segments (n) were calculated to be 90 by taking $\Delta$T = 10 K (n = $\dfrac{T_{h}-T_{c}}{\Delta T}$). For the first segment, using the value of $\Delta T$ = 10 K, $\overline{\rm T}$ = 305 K and Z$\overline{\rm T}$ = 0.0556 in Eqn 4, we get $\eta_{1}$ = 0.00044. Similarly, we can calculate the efficiency of every consecutive segment up to $\eta_{90}$ = 0.00106. By using these segmented efficiencies, we can calculate the overall efficiency by using the following formula
\begin{equation}
\eta_{overall} = 1-(1-\eta_{1})(1-\eta_{2})........(1-\eta_{n})
\end{equation}  
In order to calculate $\eta_{overall}$ at T$_{h}$ = 400 K, 500 K..... and 1200 K, we have taken the number of segments, n = $\dfrac{T_{h}-300}{10}$ = 10, 20..... and 90, respectively, whereas cold end temperatre was kept fixed at 300 K. From Fig. 11, the maximum efficiency is found to be $\sim$ 6$\%$ at T$_{h}$ = 1200 K for p-type Na$_{0.74}$CoO$_{2}$. Similarly, the maximum possible efficiencies of n-type Na$_{0.74}$CoO$_{2}$ are calculated using Eqn. 5 which is also shown in Fig. 11. The maximum efficiency observed at 1200 K is $\sim$ 16$\%$. The average of maximum efficiency of TEG made by p and n-type Na$_{0.74}$CoO$_{2}$ is found to be $\sim$ 11$\%$ at T$_{h}$ = 1200 K. This value of efficiency is more than the value of TEG made by Bi$_{2}$Te$_{3}$, which is commercially used in low temperature region.\cite{bite} Therefore, on the basis of studied ZT and efficiency, we propose that Na$_{0.74}$CoO$_{2}$ compound is an efficient TE material. Here, it is important to note that we have calculated the TE properties up to 1200 K and at this temperature one may question about the decomposition of the compound. Since, this compound is prepared through solid state reaction which is based on the diffusion process and the final sintering temperature of this compound was $\sim$ 1100 K. Thus, one can expect that the decomposition temperature of this compound will be higher than the temperature range studied in this work. Thus, a rigorous effort should be done in synthesizing the n-doped Na$_{x}$CoO$_{2}$ compound. 

\section{Conclusions}
In conclusion, the temperature dependence of TE properties (S and $\kappa$) of Na$_{0.74}$CoO$_{2}$ compound was measured in the temperature region $300-620$ K. The observed values of S were found to be $\sim$ $64-118$ $\mu$V/K in the temperature region $300-620$ K. The positive values of S indicate the dominating p-type behaviour of the compound. Low values of $\kappa$ ($2.2-2.6$ W/m-K) are observed for this compound which help to increase the ZT value. In order to understand the experimental results, ground state electronic structure calculations were carried out using spin-polarized and spin-unpolarized DFT and DFT + \textit{U} methods. Spin-polarized DFT + \textit{U} calculation is giving the best match with experiment when the value of \textit{U} and $\mu$ were chosen to be 4 eV and 220 meV, respectively. Then, we have estimated the ZT value up to 1200 K on the basis of computational understanding. The maximum value of ZT is found to be $\sim$ 0.67 at 1200 K. Computational results suggest to find out the n-type behaviour of the compound which leads to give the high ZT value of $\sim$ 2.7 at 1200 K with electron doping of $\sim$ 5.1$\times$10$^{20}$ cm$^{-3}$. Finally, the maximum possible values of $\eta$ of TEG made by p and n-type Na$_{0.74}$CoO$_{2}$ were calculated as a function of hot end temperature. The maximum possible value of efficiency is found to be $\sim$ 11 $\%$ by fixing the cold and hot end temperature at 300 K and 1200 K, respectively. The calculated values of ZT and efficiency suggest to use this compound as a promising TE candidate for high temperature applications.


\end{document}